\documentclass[10pt,journal,final,twocolumn]{IEEEtran}

\IEEEoverridecommandlockouts
\usepackage{cite}
\usepackage{amsmath,amsthm,amssymb,amsfonts,braket}
\usepackage{algorithmicx}
\usepackage{graphicx}
\usepackage{textcomp}
\usepackage{xcolor}
\usepackage{algorithm,balance}
\usepackage[noend]{algpseudocode}
\usepackage{breqn}
\usepackage{physics}
\usepackage{subfigure}
\usepackage{xcolor}
\usepackage{mathtools}

\usepackage{epsfig}

\newcommand{\Figdir}{./}

\newcommand{\singlefig}[3]{
\begin{figure}
\centerline{
    \setlength{\epsfysize}{0.23\textwidth}
 \epsffile{\Figdir#1}
} \vspace{-0.1in}\caption{#2} \label{fig:#3}
\end{figure}
}

\newcommand{\tydubfigsingle}[6]{
\begin{figure}
\centerline{
    \begin{minipage}{0.30\textwidth}
      \begin{center}
        \leavevmode
        \setlength{\epsfxsize}{0.85\textwidth}
        \setlength{\epsfysize}{0.72\textwidth}
        \epsffile{\Figdir#1}
       \newline{\small (a) #2}
      \end{center}
    \end{minipage}
    \hspace*{-1cm}
    \begin{minipage}{0.30\textwidth}
      \begin{center}
        \leavevmode
        \setlength{\epsfxsize}{0.85\textwidth}
        \setlength{\epsfysize}{0.72\textwidth}
        \epsffile{\Figdir#3}
       \newline{\small (b) #4}
      \end{center}
    \end{minipage}
} \caption{#5}\label{fig:#6}
\end{figure}
}

\def\BibTeX{{\rm B\kern-.05em{\sc i\kern-.025em b}\kern-.08em
    T\kern-.1667em\lower.7ex\hbox{E}\kern-.125emX}}

\makeatletter
\newcommand{\linebreakand}{%
  \end{@IEEEauthorhalign}
  \hfill\mbox{}\par
  \mbox{}\hfill\begin{@IEEEauthorhalign}
}
\makeatother

\begin{document}

\title{Blockwise Post-processing in Satellite-based Quantum Key Distribution}

\author{Minu J. Bae, Nitish K. Panigrahy, Prajit Dhara,~\IEEEmembership{Student Member,~IEEE,} Md Zakir Hossain, Walter O. Krawec, 
	Alexander Russell, Don~Towsley,~\IEEEmembership{Fellow,~IEEE,}
	Bing~Wang,~\IEEEmembership{Senior Member,~IEEE} 
 \IEEEcompsocitemizethanks{Manuscript received xx xx, 2024. M. Bae, Md. Z. Hossain, W. Krawec, A. Russell, and B. Wang are with the University of Connecticut. N. Panigrahy is with 
 Binghamton University. D. Towsley is with the University of Massachusetts, Amherst. P. Dhara is with the University of Arizona. This research was supported in part by the NSF grant CNS-2402861, NSF-ERC Center for Quantum Networks grant EEC-1941583,  MURI ARO Grant
W911NF2110325, and NSF CCF-2143644. Any opinions, findings, and conclusions or recommendations expressed in this material are those of the authors and do not necessarily reflect the views of the funding agencies. }
}

\maketitle

\begin{abstract} 
Free-space satellite communication has significantly lower photon loss than  terrestrial communication via optical fibers. Satellite-based quantum key distribution (QKD) leverages this advantage and provides a promising direction in achieving long-distance QKD. While the technological feasibility of satellite-based QKD has been demonstrated experimentally, optimizing the key rate remains a significant challenge. In this paper, we argue that improving classical post-processing is an important direction in increasing key rate in satellite-based QKD, while it can also be easily incorporated in existing satellite systems. In particular, we explore one direction, {\em blockwise} post-processing, to address highly dynamic satellite channel conditions due to various environmental factors. This blockwise strategy divides the raw key bits into individual blocks that have similar noise characteristics, and processes them independently, in contrast to traditional non-blockwise strategy  that treats all the raw key bits as a whole. Using a case study, 
 we discuss the choice of blocks in blockwise strategy, and show that blockwise strategy can significantly outperform non-blockwise strategy. Our study demonstrates the importance of post-processing in satellite QKD systems, and presents several open problems in this direction. 
\end{abstract}

\section{Introduction}
Quantum Key Distribution (QKD) uses the principles of quantum mechanics to enable two parties, commonly referred to as Alice and Bob, to establish a shared secret key~\cite{Pirandola20:QKD-advances}.
A fundamental characteristic of QKD is that it can detect any eavesdropping attempt, providing a level of security that is not achievable using classical cryptographic methods alone.
Despite much progress, QKD over long distance remains a challenge.
Specifically, quantum signals experience exponential loss while propagating through optical fibers, significantly restricting their travel distance. To overcome this challenge, many researchers are turning to satellite-based quantum communication,
leveraging the much lower loss in free space compared to fiber links~\cite{Chang2024:Entanglement}. While several experimental demonstrations have shown the technological feasibility of satellite-based QKD (e.g.,~\cite{wang2013direct}), much work remains in optimizing the overall QKD system performance and speed.

In general, QKD protocols operate in two distinct stages.  First is a \emph{quantum communication stage}, where users establish \emph{raw keys} using the actual quantum channel and the actual ``quantum-capable'' hardware (i.e., hardware capable of sending and measuring quantum states).  The second stage, the \emph{classical post-processing stage}, uses purely classical communication and computation to distill a secret key from the raw key data. 
The performance and efficiency of QKD can be improved in either stage. For satellite-based QKD, improvement in the second post-processing classical stage is a particularly important direction. This is because 
such changes would not require users to invest or install new quantum hardware, and hence  can be easily incorporated in existing satellite systems, even after launching the satellites.

As such, we focus on the classical post-processing stage of satellite-based QKD systems in this paper. 
An interesting challenge in satellite communication
is the highly dynamic nature of the quantum channel,
caused by various environmental circumstances (e.g., time of day, weather conditions).
Such channel dynamics can significantly affect the key rate of QKD and need to be considered carefully. For example, existing studies~\cite{Erven12:turbulent,Vallone15:ARTS,Wang18:Prefixed-threshold} show that it is beneficial to discard measurements with a high quantum error rate under dynamic channel conditions, which can improve the overall key rate.

In this paper, we present a classical post-processing approach to further improve the key rate.
In particular, we consider {\em blockwise} post-processing (referred to as key-pool segmenting in \cite{Amer24:dynamic-ICDCS}), which divides the raw key bits into multiple individual ``blocks'' based on their noise characteristics, and processes each block independently. In contrast to~\cite{Amer24:dynamic-ICDCS}, which focuses on terrestrial networks and considers only idealistic asymptotic scenarios, we use this strategy for satellite-based QKD systems in realistic {\em finite-key} scenarios.

We use a case study to investigate
the benefits of blockwise post-processing in improving key rate. In the case study, we consider a satellite-based QKD system with realistic noise and loss models, multiple satellite altitudes, and three ground station pairs.  Leveraging our finite-key results, we first identify an {\em optimal fidelity threshold} for a block of raw key bits, and show that discarding raw key bits with fidelity below this threshold leads to the best results. We then discuss the choice of blocks in a blockwise strategy, and show that a blockwise strategy leads to more secret key bits than a non-blockwise strategy.
The improvement is up to 8.6\% considering one day of data.  
Our results highlight the importance of 
designing classical post-processing techniques to improve the key rate in satellite-based QKD systems. 

The rest of the paper is organized as follows. In Section~\ref{sec:sat-QKD}, we present background on satellite-based QKD. In Section~\ref{sec:blockwise}, we describe the blockwise strategy and the key rate in finite-key scenarios. The benefits of using this post-processing strategy in satellite-based QKD are shown in Section~\ref{sec:results}. Section~\ref{sec:others} points out other promising directions for improving key rate in satellite QKD systems. Last, Section~\ref{sec:concl} concludes the paper. 

\section{Satellite-based QKD} \label{sec:sat-QKD}

\begin{figure}[t]
    \centering  \includegraphics[width=0.48\textwidth]{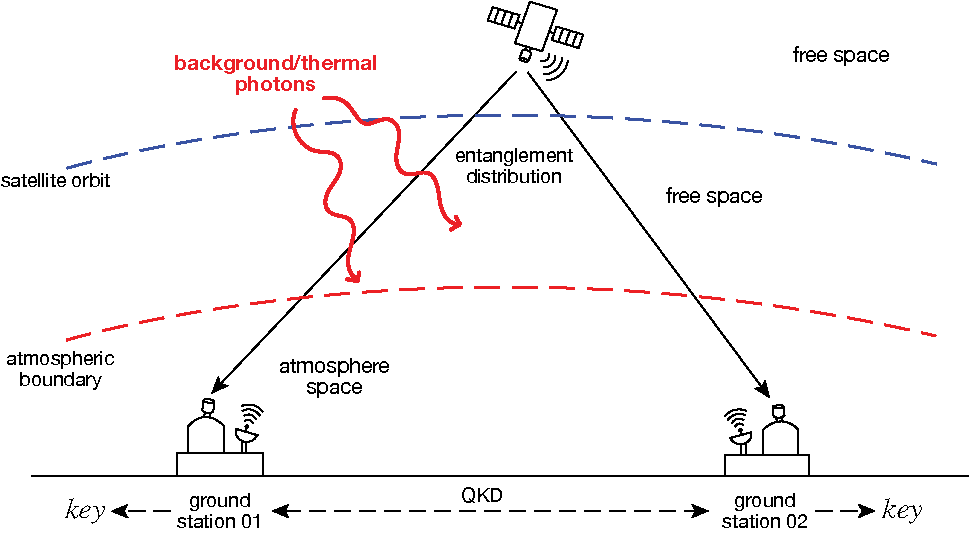}
    \caption{Dual-downlink based QKD from one satellite to two ground stations. }
    \label{fig:sat-QKD}
\end{figure}

Two common QKD approaches are  {\em prepare-and-measure} 
 and {\em entanglement-based}, exemplified by BB84~\cite{BB14:QKD} and E91~\cite{Ekert91:E91} protocols, respectively. 
 Both approaches can be used in satellite systems. In prepare-and-measure approach, Alice prepares the quantum states and sends them to Bob, which can be in the downlink direction, i.e., space-to-ground, or in the uplink direction, i.e., ground-to-space. While space-to-ground is more efficient since the atmospheric turbulence acts at the end of the optical transmission, it has the disadvantage that operating a quantum source on-board a satellite is more challenging than on the ground. 
 In an entanglement-based approach, a satellite has photon sources that generate entangled pairs, and sends them to Alice and Bob on the ground. Specifically, for each entangled pair, the satellite transmits one photon to Alice and the other to Bob, forming {\em dual-downlink} entanglement distribution  shown in Fig.~\ref{fig:sat-QKD}.
 The entanglement-based approach offers advantages over the prepare-and-measure approach in that the entanglement source (i.e., satellite) does not need to be trusted, which is important since the satellite operator can be potentially malicious. We therefore focus on the entanglement-based dual-downlink setting in the rest of the paper; the blockwise post-processing technique can also be applied to single downlink or uplink prepare-and-measure approaches.

\subsection{Dual-downlink Satellite-based QKD}
Let us consider a specific protocol, E91~\cite{Ekert91:E91}, to illustrate entanglement-based QKD in the dual-downlink setting.  This protocol, has two stages: quantum communication stage, followed by classical post-processing stage. The quantum communication stage of the protocol has the source (at the satellite) sending entangled particles to Alice and Bob (the two ground stations), one particle per round.  Alice and Bob choose, independently, to measure in one of two bases, $X$ or $Z$.  If they choose the same basis, they should receive a correlated outcome due to properties of the entangled particles.  Otherwise, if they measure in opposite bases, the result is random.  Note that the original version of the protocol assumes the basis choices are chosen with equal probability, though modern variants will bias the basis choice to improve performance.  The basis choices of Alice and Bob, but not their measurement outcomes, are broadcast over a public, authenticated, classical channel.  The two parties discard all rounds where they measured in alternative bases; this leaves Alice (respectively Bob) with classical bit strings $Z_A$ and $X_A$ (respectively $Z_B$ and $X_B$) for the measurements in $Z$ and $X$ bases, respectively.  

The classical post-processing stage involves Alice and Bob disclosing a random subset from each of their $Z$ and $X$ classical strings.  This subset allows Alice and Bob to test the correlation between their measurement outcomes in both bases.  Ideally their strings will match exactly. However, natural and adversarial noise may cause bit errors.  From this information, Alice and Bob are able to determine an upper bound on the information an adversary may have, due to various properties of quantum mechanics (including the no-cloning theorem and entropic uncertainty).  The remaining bits in both strings are combined to form their \emph{raw keys}. The raw keys are partially correlated (natural and adversarial noise may cause bit-flip errors) and partially secret (an adversary may have some partial information on the raw key), and hence must be further processed before they can be used as a secret key. Alice and Bob then run an error correction protocol (which leaks some additional information to a potential adversary) followed by a privacy amplification protocol.  The first ensures that the raw keys are identical, while the second involves hashing the raw key down to a smaller, but secret, key.  The size of the final secret key depends on how much information the adversary, Eve, is estimated to have---the more information she has on the raw key, the more privacy amplification must shrink the raw key by.  The \emph{key rate} of a QKD protocol, then, is the number of secret key bits produced after privacy amplification, divided by the total number of rounds used.

\subsection{Quantum Channel Loss and Noise} \label{sec:model}
Dual-downlink entanglement distribution 
(see Fig.~\ref{fig:sat-QKD}) relies 
on photonic entangled-pair generation sources and free-space optical communication strategies. 
Specifically, the satellite platform contains an entanglement generation source and transmission optics to transmit entangled pairs to the two ground stations, which contain receiver optics and adaptive optics (to minimize atmospheric distortion) to receive photons from the satellite. The two ground stations must be simultaneously within the field-of-view of the satellite. The overall system (satellite and ground stations) needs to have accurate time synchronization, pointing and tracking, and classical communication links for QKD to operate successfully.

Henceforth, we assume that the satellite utilizes spontaneous parametric down-conversion (SPDC) based dual-rail polarization entanglement sources that are well-studied and widely used~\cite{dhara2022heralded}.
In such entanglement sources, a two-qubit entangled Bell state requires four orthogonal modes (i.e., two pairs of modes) to encode, and takes the form  
    $\ket{\Psi^{\pm}} = \frac{1}{\sqrt{2}}(\ket{1,0;0,1}\pm\ket{0,1;1,0})$.
The vacuum state is $\ket{0,0;0,0}$. There are other spurious high-order photon states. The pump power of the source needs to be low 
so that high-order photon states are negligible.

The satellite quantum communication link  involves free-space optical (FSO) transmission, and hence the analysis of such links must account for the characteristics of the optical channel.
The transmission loss for each entangled photon scales  quadratically with free space propagation length (i.e., from the satellite to the ground) and exponentially with 
atmospheric
propagation length (i.e., from the atomospheric boundary to the ground) \cite{panigrahy2022optimal}.
FSO transmission reduces the mean photon number of the transmitted state, and can result in a pure state becoming mixed. Such effects impede the probability of successfully delivering the entangled photons to both ground stations, as well as affecting the fidelity (to the ideal entangled state) of the delivered entangled photons. Namely, it reduces the probability of receiving a perfect pair of entangled photons at the two ground stations.

In addition to transmission loss, atmospheric FSO transmission channels have to contend with a variety of noise processes. One such source of noise are the unfiltered background photons. Any excess photons in the channel will cause false events to be treated as successes, thereby impacting the fidelity of the  entangled pair that should have been delivered. The main contributor of the background photon flux is commonly associated with the brightness of the sky and varies drastically depending on the time of the day. More specifically, the level of background photon flux is at its highest during clear daylight, and at its lowest during  clear nighttime. We compute the fidelity of the generated entangled state between two ground stations by modeling the arrival of unfiltered background photons as detector dark click events.

\subsection{Improving Key Rate in Satellite-based QKD}
Real-world deployment and commercial viability of satellite-based QKD depend heavily on its key rate.
The key rate can be improved through innovative quantum technologies, 
e.g., high-performance entanglement sources, wavelength multiplexing, and high-dimensional QKD.
Another direction, which we focus on in this paper, is innovation in the classical post-processing stage, which can be designed for general quantum technologies, or customized for specific quantum hardware. 
Such post-processing innovations have the advantage that they can be easily deployed in existing systems,  even after the launch of a satellite. 

We next outline one such post-processing technique. From the above sections, it is clear that satellite QKD has several unique challenges, one of particular note being the dynamic nature of the noise in the channel due to changing environmental effects.  Recalling our discussion of the E91 protocol earlier, we see that such dynamics cause the noise level in both the $Z$ and $X$ strings (the raw keys) to vary with time.  Ordinarily, classical post-processing will treat the entire string as one homogeneous block, computing the adversary Eve's uncertainty on the entire string as a function of the average noise.  However, this gives a pessimistic key rate since, for certain, potentially large sections, of these raw keys, the noise is much lower and, thus, Eve will have potentially much less information on that particular block.  Thus, it makes sense to break the $Z$ and $X$ strings into ``blocks''---each block having an expected noise level that is homogeneous.  Error correction and privacy amplification can then be run on each block independently, assuming the blocks are of sufficient size. As we show later, this blockwise strategy can produce significant  improvement in key generation rates.

\section{Blockwise Post-processing} \label{sec:blockwise}

\begin{figure}[t]
    \centering  \includegraphics[width=0.40\textwidth, trim = 0.cm 0.5cm 0cm 2.5cm, clip]{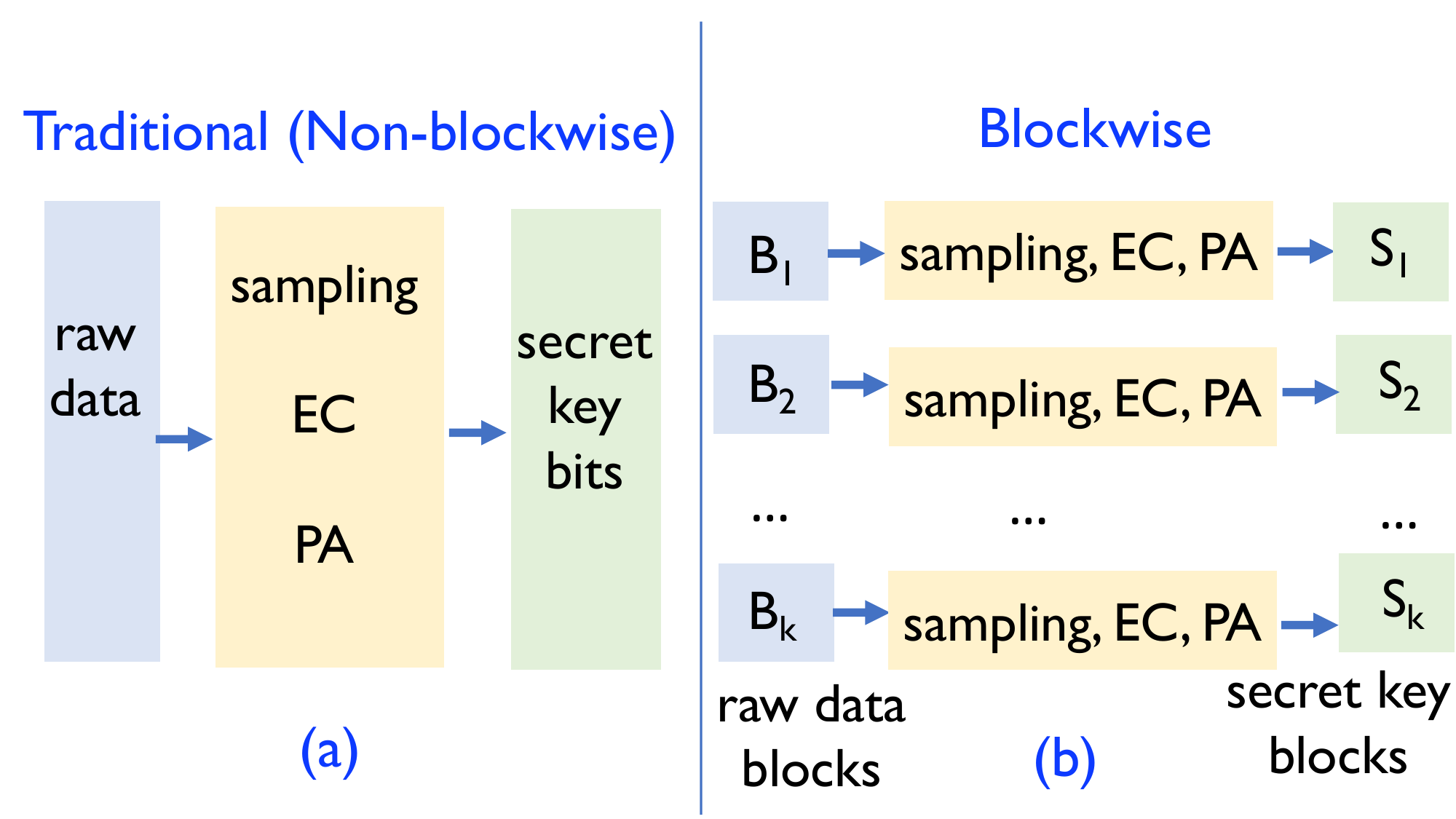}
    \vspace{-0.1in}
    \caption{Illustration of sampling, error correction (EC), and  privacy amplification (PA) in  non-blockwise and blockwise strategies. }
    \label{fig:noblock-vs-block}
\end{figure}

In this section, we present the blockwise strategy outlined earlier and analyze its key rate. As illustrated in Fig.~\ref{fig:noblock-vs-block}(b), this blockwise strategy divides the raw data after the quantum communication stage into $k$ blocks, $B_1, \ldots, B_k$, each corresponding to similar channel characteristics, especially in terms of noise. Then sampling, error correction, and privacy amplification are applied separately to each block to obtain the secret key blocks $S_1, \ldots, S_k$, which are then concatenated into a single secret key block. This is in contrast to the traditional  non-blockwise strategy shown in   Fig.~\ref{fig:noblock-vs-block}(a), whereby the raw data is treated as a single block (agnostic to the dynamics of the quantum channel), on which sampling, error correction and privacy amplification are applied. 

Before analyzing the key rate for non-blockwise and blockwise strategies, we first present the attack model. We then present the key rate analysis for these two strategies, both in the finite-key scenario. At the end, we comment on the asymptotic key rate.  

\smallskip
\noindent{\bf Attack model.} In our analysis below, we make a standard assumption on the power of the adversary, Eve: (1) she fully controls the source, and (2) she is allowed to prepare the entire signal state at once (i.e., she can prepare signals for all $M$ rounds at once in her lab) as opposed to sending signals one at a time.  We assume the parties then delay measurement until the entire signal is received.  This gives Eve greater potential information than the more practical setting where rounds are performed ``on-the-fly'' and so, in real life, the performance of the secret key generation rate can only be better.  The above is a common assumption in QKD security proofs.

In detail, we assume the adversary controls the entanglement source and is allowed to create an entangled state of $2M$ particles, sending $M$ particles to Alice and $M$ particles to Bob; Alice and Bob measure their particles in the $Z$ or $X$ basis.  Loss in the channel and basis mismatch cause only $N\le M$ particles to be useful as raw data for Alice and Bob.

\smallskip
\noindent{{\bf Non-blockwise strategy.}} In this case, suppose Alice and Bob  randomly choose $m$ samples 
from the $N$-bits of raw measurement data, $m<N/2$.  This sampling allows Alice and Bob to estimate the error rate in the entire raw key, denoted as $Q$.  
After that, the remaining $n=N-m$ raw key bits are run through an error correction process, which leaks an additional $\lambda_{EC}$ bits to the adversary. A  test is then run by hashing the error corrected raw key and testing correctness between Alice and Bob.
Finally, privacy amplification is run, outputting a secret key of size $\ell \le n$. It is guaranteed that the final joint state will be $\epsilon_{sec}$-close (in trace distance, for any $\epsilon_{sec} > 0$) to an ideal secrete key, namely one that is uniform random and independent of Eve's ancilla \cite{renner2008security}.
That is, the final secret key system will be $\epsilon_{sec}$ close (in trace distance) to a truly uniform random key, $I/2^\ell$, which is also completely independent of Eve's system.

Using results in \cite{tomamichel2012tight}, the non-blockwise case can be shown to have an overall secret key length of
\begin{equation} \label{eq:keyrate-nonblock}
    \ell_{\text{non-block}} = n(1-h(Q+\mu)) - \lambda_{EC} - \log\frac{2}{\epsilon_{sec}^2\epsilon_{cor}}\,,
\end{equation}
where $\epsilon_{cor}>0$ is another user-specified  parameter determining the failure rate of the correctness portion of the protocol (i.e., Alice and Bob will have the same secret key, except with probability at most $\epsilon_{cor}$).
Above, $\lambda_{EC}$ represents the information leaked during error correction. In upcoming evaluations,
we simply set $\lambda_{EC} = nh(Q+\mu)$, where $h(x)=-x\log x - (1-x)\log(1-x)$ is the binary entropy function, and $\mu$ is a result of finite sampling effects. Following standard classical sampling arguments \cite{tomamichel2012tight}, $\mu$ is set to 
\begin{equation} \label{eq:mu}
\mu = \sqrt{\frac{(n+m)(m+1)}{nm^2}\ln\frac{2}{\epsilon_{sec}}}.
\end{equation}

In Eq. (\ref{eq:keyrate-nonblock}), to optimize the key rate, one needs to determine the optimal sample size, i.e., the optimal value of $m$ given $N$. We explore doing this numerically in Section~\ref{sec:results}.   

\smallskip
\noindent{{\bf Blockwise strategy.}} In the blockwise case, the setting is similar and we may again use results from \cite{tomamichel2012tight} to distill each sub-block into secret keys independently.
Here, let $B_i$ be the size of the $i$'th block.
Now, a random subset of size $m_i$ for each block $B_i$ is chosen, which is used to determine the error rate, denoted $Q_i$, in this individual block.  
Error correction, a correctness test, and finally privacy amplification are then performed individually on each block.  By treating each block independently, one may derive an additive structure on the final key size using \cite{tomamichel2012tight}.  That is, we can compute the secret key size of block $i$ using the same technique used to derive Eq. (\ref{eq:keyrate-nonblock}).  This leads to a block key size of
\[
\ell_i = (B_i-m_i)(1-h(Q_i+\mu_i)) - \lambda_{EC}^{(i)} - \log\frac{2}{\epsilon_{sec}^2\epsilon_{cor}},
\]
where the value of $\mu_i$ is identical to $\mu$ in Eq. (\ref{eq:mu}), except replacing $m$ with $m_i$ and $n$ with $B_i-m_i$, and $\lambda_{EC}^{(i)}$ is the amount of information leaked during error correction of block $i$.  In our evaluations, we set  $\lambda_{EC}^{(i)} = (B_i-m_i)h(Q_i+\mu_i)$.

With $\ell_i$ as above, the total secret key size for the blockwise strategy is $\ell_{\text{block}} =\sum_{i=1}^k \ell_i$,
 where $k$ is the total number of blocks.  Note that the above equation forces Alice and Bob to have a sufficient amount of raw key material in each individual block, otherwise $m_i$ will be too small, causing $\mu_i$ to be large, forcing the key-size $\ell_i$ to go to zero.  Thus, this can only lead to a non-zero key for block $i$ if there is sufficient material in block $i$ to sample from.

\smallskip
\noindent{{\bf Asymptotic key rate.}} To determine theoretical upper-bounds, we also consider the asymptotic scenario, where the number of signals approaches infinity.  In this instance,  the key rate for the non-blockwise strategy is simply $1-2h(Q)$, while the key rate for the blockwise strategy converges to $\sum_i p_i(1-2h(Q_i))$, where $p_i$ is the proportion of total raw key bits used in block $i$ as the size of the raw key approaches infinity.

The above equations give immediate intuition as to why blockwise processing can lead to higher key rates.  For non-blockwise, the total error $Q$ is actually the average error over all individual blocks.  Due to the concavity of Shannon entropy, the key rate of blockwise strategy is no less than the non-blockwise strategy in the asymptotic scenario. 

\section{Benefits of Blockwise Post-processing}\label{sec:results}

\tydubfigsingle{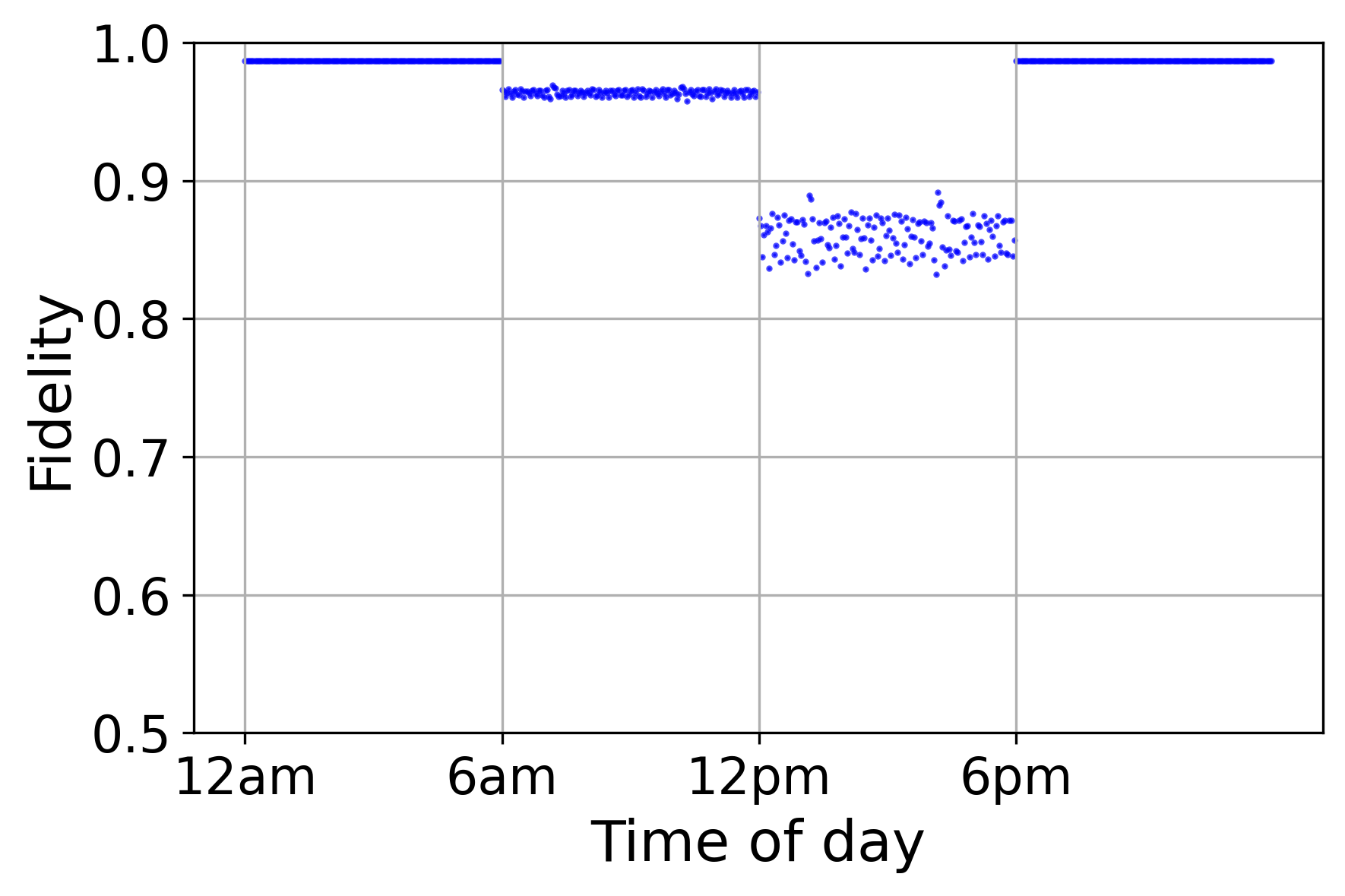}{$A=500$ km.}{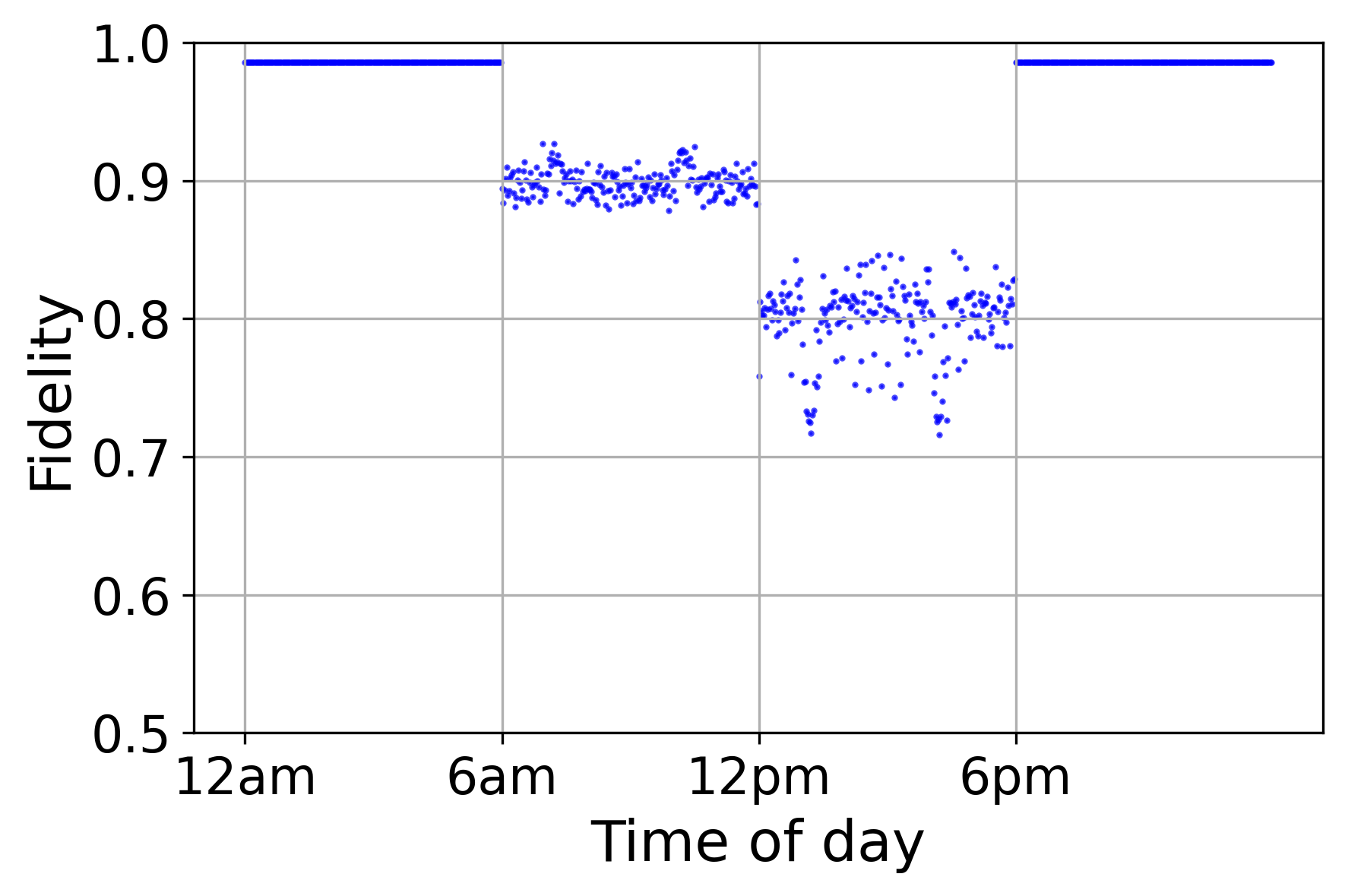}{$A=1300$ km.}{Fidelity over time for Toronto-DC pair. For clarity, each point in the figure represents the average fidelity over one minute.}{fidelity}

\singlefig{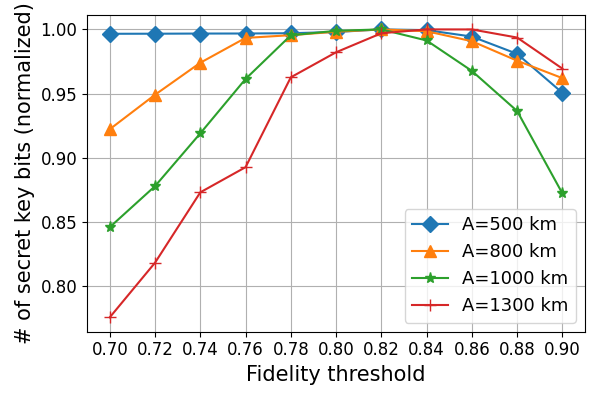}{Normalized number of secret key bits versus fidelity threshold, Toronto-DC, non-blockwise strategy.
For each altitude, the number of secret key bits is normalized by the maximum number of secret key bits (obtained using the optimal fidelity threshold) for that altitude. }{threshold}

\singlefig{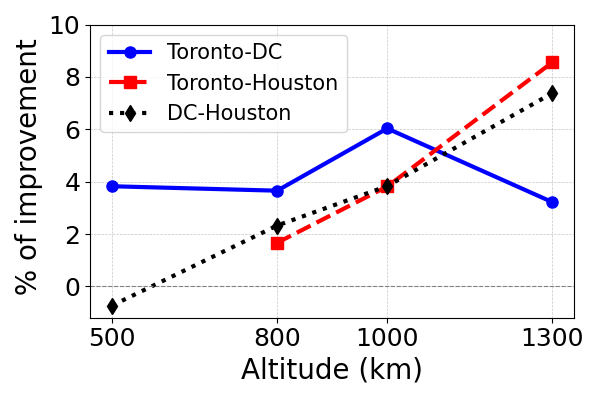}{Improvement of  
blockwise strategy over non-blockwise strategy.
No key is generated for Toronto-Houston when the altitude is 500 km. }{improve}

We next use a case study to demonstrate the benefits of blockwise post-processing strategy. In this case study, 
we consider a polar constellation of low-earth-orbit (LEO) satellites consisting of 20
rings, each ring with 20 satellites. These satellites orbit around the Earth at the same altitude $A$, which is set to 500, 800, 1000, or 1300 km. We consider three
ground stations in North America: 
Washington D.C. (DC), Toronto, and Houston, forming a total
of three ground station pairs. 
Each satellite is equipped with an SPDC entanglement source (see Section~\ref{sec:model}) that generates $10^9$ entangled photons per second. The pump power is set to a low value of 0.01 so that high-order photon contributions are negligible. For  successful delivery of an entangled pair, the elevation angles between the satellite and the two ground stations need to exceed an angle, which is set to 20$^{\circ}$ in our simulation. When a ground station pair can be served by multiple satellites, we select the one with the highest estimated fidelity to the ground station pair.

To account for the time-of-day effects, as in \cite{Maule2024:scheduling}, we consider four time points, 12am, 6am, 12pm, and 6pm, in a day. For simplicity, we divide the entire day into four 6-hour intervals, starting respectively from the above four time points, 
and assume the same solar radiance
for each interval. The solar radiance for each of the four time points is based on the measurements for March 15, 2022~\cite{Maule2024:scheduling}.

The key rate for blockwise and non-blockwise strategies uses the finite-key results in Section~\ref{sec:blockwise}. It requires determining an optimal sampling rate, which is determined numerically.  
Specifically, for non-blockwise strategy, the sampling rate (i.e., $m/N$) is varied from $10^{-5}$ to $0.05$, 
and an optimal value is selected using grid search to maximize the key rate. For blockwise strategy, the optimal sampling rate for each block (i.e., $m_i/B_i$) is determined separately using grid search. 

\smallskip
\noindent{\bf Channel dynamics.} Using the quantum channel noise model, we obtain the fidelity per second (when there is at least one satellite in view of the ground station pairs).  Fig.~\ref{fig:fidelity} plots fidelity for Toronto-DC over time for two altitudes, $A=500$ km and 1300 km.
 We see that fidelity changes dynamically, roughly following four 6-hour intervals as in our assumption. 
 The first and last intervals, from 12am to 6am and from 6pm to 12am, have the highest fidelity, followed by the second interval, 6am to 12pm;  the third interval, from 12pm to 6pm, has the lowest fidelity. 
 In addition, fidelity also change within each interval due to changing satellite distance and angle from the ground stations.

Existing studies~\cite{Erven12:turbulent,Vallone15:ARTS,Wang18:Prefixed-threshold} propose discarding measurements that have low fidelity. While it is clear that measurements with very low fidelity (e.g., significantly lower than the noise tolerance of the QKD scheme) should be discarded, it is not clear at exactly which fidelity level should the measurements be discarded for a given setting. We next investigate the choice of {\em fidelity threshold}, i.e., below which fidelity value should the raw key bits be discarded.  
Eq. (\ref{eq:keyrate-nonblock}) shows the opposing impacts of 
discarding low-fidelity raw key bits on the total number of secret key bits: on the one hand, it leads to lower error rate ($Q$); on the other hand, it leads to fewer raw key bits ($n$) and potentially higher sampling impact ($\mu$). As a result, we expect an {\em optimal fidelity threshold}: discarding raw key bits with fidelity below this threshold leads to the maximum number of secret key bits.

Fig.~\ref{fig:threshold} plots the normalized number of secret key bits when varying the fidelity threshold from 0.70 to 0.90 for Toronto-DC, considering all the raw key bits generated in a day. The results for altitude $A=500$, 800, 1000, and 1300 km are plotted in the figure. We see that there indeed exists an optimal fidelity threshold for each setting, and the optimal values are similar (0.82 or 0.84) across different settings.    
Similarly, we determine the optimal fidelity thresholds for the other two ground station pairs across various altitudes.  The results presented below for each setting are obtained using the optimal fidelity threshold for that setting. In blockwise strategy, we identify optimal fidelity threshold for each block separately.

In the above, we have demonstrated the impact of fidelity threshold and identified an optimal fidelity threshold based on the fidelity of the raw key bits for each  setting. In practice, the fidelity of the raw key bits can be estimated by sampling a small number of raw bits in a short time window or using other methods (e.g.,  characterizing free-space optical links, measuring signal-to-noise ratio), which is left as future work.

\smallskip
\noindent{\bf Benefits of blockwise strategy.} We next compare the number of secret key bits under non-blockwise and blockwise strategies.
For blockwise strategy, we explore two methods: (i) 2-block, the first with very high fidelity ($\ge 0.98$), and the second containing the rest, and (ii) 3-block, the first with fidelity $\ge 0.98$, the second with fidelity in $(0.90,0.98)$, and the third containing the rest. From Fig.~\ref{fig:fidelity}, we see three distinct fidelity ranges, indicating that 3-block might be a good choice. On the other hand, we also need to consider the size of a block, since sampling effect is higher for smaller blocks (see Eq. (\ref{eq:mu})). Of all the settings we explore, we observe 3-block outperforms 2-block strategy in most cases. However, in some cases (e.g., Toronto-Houston, $A=800$ and 1000 km), 2-block outperforms 3-block strategy. For a given setting, finding the best blockwise strategy is an open problem that is left as future work.

Fig.~\ref{fig:improve} plots the percentage of improvement of the blockwise strategy over
the non-blockwise strategy, i.e.,  $(\ell_{\text{block}}-\ell_{\text{non-block}})/\ell_{\text{non-block}}$, where $\ell_{\text{block}}$ and $\ell_{\text{non-block}}$ denote the total number of secret key bits under blockwise (2-block or 3-block, whichever is better) and non-blockwise strategies, respectively. 
We see that blockwise strategy leads to up to $8.6\%$ 
more key bits than non-blockwise strategy. Only in one case (DC-Houston, $A=500$ km), the blockwise strategy leads to fewer keys than the non-blockwise strategy due to the small number of entanglements that are successfully received, and hence suffers from higher sampling rate; see Eq. (\ref{eq:mu}). On the other hand, when applied to 10 days of raw key bits, blockwise strategy leads to more keys than non-blockwise strategy in all settings (figure omitted), and the improvement is larger than when considering one day.  

\section{Open Problems} \label{sec:others}

Our case study in Section~\ref{sec:results} demonstrates the benefits of applying blockwise strategy 
in satellite-based QKD, and in general, the importance of carefully considering the post-processing stage to improve key rate. In the following, we discuss several open problems related to improving key rates in satellite QKD systems.  

\smallskip
\noindent{\bf Real-time dynamic channel conditions.} In addition to time-of-day effects, many other environmental factors (e.g., weather, cloud coverage) can lead to dynamic quantum channel conditions. Often times, real-time monitoring is needed to identify such dynamic conditions, estimate quantum error rates, and then divide the raw data into blocks accordingly. In addition, one needs to determine what range of error rates fall into one block so that the size of each block is sufficiently large, while the error rates of a block is not too heterogeneous.

\smallskip
\noindent{\bf  Other components in post-processing.} 
In our blockwise strategy, after raw data are divided into blocks, error correction and privacy amplification are run for each block separately. Another approach is to bound the quantum min entropy of each block in the blockwise strategy directly, and run a single privacy amplification processes over all the blocks. That is, use a single invocation of privacy amplification, as in the non-blockwise strategy, yet still retain the benefit of increased key lengths as in blockwise post-processing. In general, the multiple components in post-processing can be jointly considered for better efficiency.

\smallskip
\noindent{\bf Connecting quantum and classical techniques.} 
While we emphasize the importance of classical post-processing  in improving key rate in satellite-based QKD, which is in parallel to quantum techniques, another direction is connecting quantum and classical techniques. For instance, for satellite systems that can tune quantum parameters (e.g., pump power, beam width), an interesting direction is to combine the choice of the parameters with the post-processing technique, and determine them jointly to optimize key rate. In addition, customized post-processing techniques can be developed for certain new quantum technologies. One example is high-dimensional QKD, which features superpositions of many (instead of two) orthogonal quantum states. This is regarded as a promising technology for satellite-based QKD since it is expected to exhibit increased channel capacity and resilience than two-dimensional systems. For such high-dimensional systems, the error characteristics of the quantum channel may differ from those for two-dimensional systems, and hence specialized post-processing techniques may benefit such systems.

\smallskip
\noindent{\bf Other attack models.}  In Section~\ref{sec:blockwise}, we consider a standard attack model where the adversary has full control of the quantum channels from the satellite to the two ground stations. It would be interesting to consider alternative, more realistic security models (in terms of adversarial capabilities), 
and how post-processing methods can help in this context.  
For instance, it is likely Eve will not control everything and would have to launch an airborne object (e.g., a drone) to intercept incoming signals from a satellite. However, such an object would be unable to capture everything.  This was a security model introduced in \cite{ghalaii2023satellite}. How blockwise processing, or other alternative methods, can benefit key rates in this scenario is worthy of investigation.

\section{Conclusion} \label{sec:concl}
In this paper, we pointed out that improving classical post-processing in satellite-based QKD is an important direction in improving key rate. 
In particular, we explored one direction, {\em blockwise} post-processing.
Using a case study, we showed that the blockwise strategy can lead to significantly higher key rates than the traditional non-blockwise strategy that is agnostic to the dynamics of the quantum channel. We also pointed out open problems that can further improve key rates of satellite-based QKD.


\begin{thebibliography}{10}

\bibitem{Pirandola20:QKD-advances}
S.~Pirandola, U.~L. Andersen, L.~Banchi, M.~Berta, D.~Bunandar, R.~Colbeck, D.~Englund, T.~Gehring, C.~Lupo, C.~Ottaviani, J.~L. Pereira, M.~Razavi, J.~S. Shaari, M.~Tomamichel, V.~C. Usenko, G.~Vallone, P.~Villoresi, and P.~Wallden, ``Advances in quantum cryptography,'' {\em Advances in Optics and Photonics}, vol.~12, no.~4, 2020.

\bibitem{Chang2024:Entanglement}
A.~Chang, Y.~Wan, G.~Xue, and A.~Sen, ``Entanglement distribution in satellite-based dynamic quantum networks,'' {\em IEEE Network}, vol.~38, pp.~79--86, Jan./Feb. 2024.

\bibitem{wang2013direct}
J.-Y. Wang, B.~Yang, S.-K. Liao, L.~Zhang, Q.~Shen, X.-F. Hu, J.-C. Wu, S.-J. Yang, H.~Jiang, Y.-L. Tang, {\em et~al.}, ``Direct and full-scale experimental verifications towards ground--satellite quantum key distribution,'' {\em Nature Photonics}, vol.~7, no.~5, pp.~387--393, 2013.


\bibitem{Erven12:turbulent}
C.~Erven, B.~Heim, E.~Meyer-Scott, J.~P. Bourgoin, R.~Laflamme, G.~Weihs, and T.~Jennewein, ``Studying free-space transmission statistics and improving free-space quantum key distribution in the turbulent atmosphere,'' {\em New Journal of Physics}, vol.~14, 2012.

\bibitem{Vallone15:ARTS}
G.~Vallone, D.~Marangon, M.~Canale, I.~Savorgnan, D.~Bacco, M.~Barbieri, S.~Calimani, C.~Barbieri, N.~Laurenti, and P.~Villoresi, ``Adaptive real time selection for quantum key distribution in lossy and turbulent free-space channels,'' {\em Phys. Rev. A}, vol.~91, 2015.

\bibitem{Wang18:Prefixed-threshold}
W.~Wang, F.~Xu, and H.-K. Lo, ``Prefixed-threshold real-time selection method in free-space quantum key distribution,'' {\em Phys. Rev. A}, vol.~97, 2018.

\bibitem{Amer24:dynamic-ICDCS}
O.~Amer, W.~O. Krawec, M.~Z. Hossain, V.~U. Manfredi, and B.~Wang, ``Dynamic routing and post-processing strategies for hybrid quantum key distribution networks,'' in {\em IEEE ICDCS}, 2024.

\bibitem{BB14:QKD}
C.~H. Bennett and G.~Brassard, ``Quantum cryptography: Public key distribution and coin tossing,'' {\em Theoretical Computer Science}, vol.~560, pp.~7--11, December 2014.

\bibitem{Ekert91:E91}
A.~K. Ekert, ``{Quantum cryptography based on {Bell's} theorem},'' {\em Physical Review Letters}, vol.~67, pp.~661--663, August 1991.

\bibitem{dhara2022heralded}
P.~Dhara, S.~J. Johnson, C.~N. Gagatsos, P.~G. Kwiat, and S.~Guha, ``Heralded multiplexed high-efficiency cascaded source of dual-rail entangled photon pairs using spontaneous parametric down-conversion,'' {\em Physical Review Applied}, vol.~17, no.~3, p.~034071, 2022.

\bibitem{panigrahy2022optimal}
N.~K. Panigrahy, P.~Dhara, D.~Towsley, S.~Guha, and L.~Tassiulas, ``Optimal entanglement distribution using satellite based quantum networks,'' in {\em IEEE INFOCOM Workshops}, 2022.

\bibitem{renner2008security}
R.~Renner, ``Security of quantum key distribution,'' {\em International Journal of Quantum Information}, vol.~6, no.~01, pp.~1--127, 2008.

\bibitem{tomamichel2012tight}
M.~Tomamichel, C.~C.~W. Lim, N.~Gisin, and R.~Renner, ``Tight finite-key analysis for quantum cryptography,'' {\em Nature communications}, vol.~3, no.~1, pp.~1--6, 2012.


\bibitem{Maule2024:scheduling}
R.~Maule, N.~K. Panigrahy, N.~L. Anipeddi, P.~Dhara, D.~Kilbane, M.~Z. Hossain, W.~O. Krawec, D.~Towsley, and B.~Wang, ``Fair and efficient scheduling strategies for satellite assisted quantum key distribution systems,'' {\em Proc. of QCE}, 2024.

\bibitem{ghalaii2023satellite}
M.~Ghalaii, S.~Bahrani, C.~Liorni, F.~Grasselli, H.~Kampermann, L.~Wooltorton, R.~Kumar, S.~Pirandola, T.~P. Spiller, A.~Ling, {\em et~al.}, ``Satellite-based quantum key distribution in the presence of bypass channels,'' {\em PRX Quantum}, vol.~4, no.~4, p.~040320, 2023.

\end{thebibliography}
\end{document}